\documentclass[10pt, letterpaper]{article}
\usepackage[left=2.00cm, right=2.00cm, top=2.00cm, bottom=2.00cm]{geometry}
\usepackage{graphicx} % Required for inserting images
\graphicspath{ {./images/} }
\usepackage{amsmath}
\usepackage{authblk}
\usepackage{siunitx}
\usepackage{caption}
\usepackage{cite}
\usepackage{notoccite}
\usepackage{xurl}
\usepackage{achemso}
\setcitestyle{square,numbers}
\usepackage{pdfpages}

\title{\vspace{-2.0cm}\textbf{Adsorption of Guanidinium Cations to the Air-Water Interface}}
\usepackage{multicol}
\renewcommand{\abstractname}{} 
\renewenvironment{abstract}
 {\small
  \begin{center}
  \bfseries \abstractname\vspace{-.5em}\vspace{0pt}
  \end{center}
  \list{}{
    \setlength{\leftmargin}{1.5cm}%
    \setlength{\rightmargin}{\leftmargin}%
  }%
  \item\relax}
 {\endlist}

\author[1,2]{Franky Bernal}
\author[2]{Amro Dodin}
\author[1]{Constantine Kyprianou}
\author[1,2,3,4,*]{David T. Limmer}
\author[1,2,*]{Richard J. Saykally}
\affil[1]{Department of Chemistry, University of California, Berkeley, CA 94720, USA}
\affil[2]{Chemical Sciences Division, Lawrence Berkeley National Laboratory, Berkeley, CA 94720, USA}
\affil[3]{Materials Science Division, Lawrence Berkeley National Laboratory, Berkeley, California 94720, USA}
\affil[4]{Kavli Energy NanoScience Institute, Berkeley, California 94720, USA}

\date{}
\newenvironment{Figure}
  {\par\medskip\noindent\minipage{\linewidth}}
  {\endminipage\par\medskip}
\begin{document}

\maketitle

\begin{abstract}
\vspace*{-4em}
Combining Deep-UV second harmonic generation spectroscopy with molecular simulations, we confirm and quantify the specific adsorption of guanidinium cations to the air-water interface. Using a Langmuir analysis and measurements at multiple concentrations, we extract the Gibbs free energy of adsorption, finding it larger than typical thermal energies.  Molecular simulations clarify the role of polarizability in tuning the thermodynamics of adsorption, and establish the preferential parallel alignment of guanidinium at the air-water interface. Guanidinium is the first polyatomic cation proven to exhibit a propensity for the air-water interface. As such, these results expand on the growing body of work on specific ion adsorption.  
\end{abstract}

\begin{center}
\end{center}

\begin{multicols}{2}
Chemistry at aqueous interfaces underlies vital applications ranging from electrochemistry and catalysis to biological membrane processes and atmospheric aerosol reactions\cite{Ruiz-Lopez2020Molecular,Law2022Design,Limmer2024Molecular}. Characterizing the detailed behavior of ions at these interfaces is thus critical in advancing our fundamental understanding of many phenomena. The study of ions at the air-water interface has undergone dramatic evolution since the early theoretical work of Onsager and Samaras exploited electrostatic arguments to describe the air-water interface as being devoid of ions \cite{Onsager1934Surface}. Under classical electrostatic theory, the boundary between two dielectric media of dissimilar permittivity (e.g. air-water) engenders an “image charge repulsion” for solvated ions from the medium of lowest dielectric constant, therefore excluding all ions from the first few outermost water layers of an air-water or hydrocarbon-water interface \cite{Wagner1924surface}. However, recent studies have challenged the completeness of this description with the observation of  several anions at the air-water interface and have rigorously established the mechanistic details driving  this ion adsorption \cite{Jungwirth2000Surface,Otten2012Elucidating,Geissler2013Water,Devlin2023Agglomeration}. When an ion comes within a few water layers of the interface, the continuum picture of Onsager and Samaras begins to break down. At these length scales, the thermodynamics of ion adsorption increasingly reflects the molecular rearrangements of the solvent required to accommodate it rather than the polarization of a continuous dielectric medium. Surface active anions, which can be bulky, and highly polarizable, with solvation cavities that do not fit naturally into bulk water structure, are more easily accommodated at the interface than in the bulk. While most cations do not share these properties, and therefore do not adsorb to the interface, we show by establishing the surface activity of the guanidinium cation, that the same molecular picture of interfacial anions can also lead to cation adsorption.\par
Nonlinear spectroscopic techniques, e.g., second harmonic generation (SHG) and sum frequency generation (SFG), have emerged as powerful tools for studying interfacial ions, given their inherent high surface specificity \cite{Shen1989Optical,Yamaguchi2004Precise}. Such techniques have now been used to establish the interfacial presence of anions such as I$^-$, N$_3$$^-$, and SCN$^-$ \cite{Petersen2006Probing,Petersen2004Confirmation,Petersen2005Enhanced,Bhattacharyya2020New,Mizuno2018Charge-Transfer-to-Solvent,Viswanath2007Oriented,Devlin2022Characterizing} in both aqueous solutions and hydrocarbon solutions. It has been generally concluded that large, singly charged, weakly hydrated, and highly polarizable (“soft”) anions exhibit surface enrichment at the air-water interface. However, recent studies have prompted a reassessment of this simple picture. X-ray photoelectron spectroscopy (XPS) and SHG results have shown that the doubly-charged CO$_3$$^{2-}$ ion exhibits a strongly increased propensity for the air-water interface relative to singly-charged HCO$_3$$^-$ \cite{Devlin2023Agglomeration,Lam2017Reversed}, and this is supported and explained by theory. Cation co-solutes are typically less polarizable, better solvated, and thus are expected to be excluded from the interface \cite{Petersen2006ON}. However, XPS measurements on LiI solution liquid jets by Perrine et al. uncovered a surprising surface activity of the Li$^+$ cation, a behavior that was not found for K$^+$ in KI solutions \cite{Perrine2017Specific}. These studies find both iodide and lithium ions to be surface-enhanced relative to the bulk concentrations. Ng et al. used UV-VIS SHG measurements to study various ferric chloride complexes at the air-water interface \cite{Ng2022Iron(III)}. These authors use symmetry and resonance arguments to attribute their SHG signal to the neutral complexes [FeCl$_3$(H$_2$O)$_x$], residing at the interface. These results highlight the fact that our understanding of ions at the air-water interface is still evolving.  In the present study, we provide both theoretical and nonlinear spectroscopic evidence for an important surface-active molecular cation.\par
We consider the guanidinium cation, (Gdm$^+$), a powerful protein denaturant widely employed in protein stability studies \cite{England2011Role}. Several theoretical reports have investigated its surface activity, given its position within the Hofmeister series, as it exhibits similarities to well-known surface-active anions \cite{Vazdar2011Like-Charge,Wernersson2011Orientational,Boudon1990Monte,Houriez2017Solvation,Ekholm2018Anomalous,Werner2014Surface,Vazdar2018Arginine}. XPS measurements on aqueous guanidinium chloride and ammonium chloride solutions found a greater interfacial population of Gdm$^+$ relative to NH$_4^+$, indicated by photoemission signals more than four times higher for Gdm$^+$ \cite{Werner2014Surface}. While XPS provides atom specificity, the technique itself is ambiguous regarding probe depth. Here we employ interface-specific SHG spectroscopy in the deep-UV (DUV-SHG) to directly probe Gdm$^+$ at the air-water interface, sampling solutions of guanidinium chloride (GdmCl). This enables us to minimize signal contribution from ions in the bulk, and by studying a series of concentrations, we can extract thermodynamic information, as we have done for anions. These experiments are highly sensitive and allow us to track small changes in SHG intensity with respect to bulk ion concentration, allowing us to fit SHG intensity to a Langmuir model and extract the Gibbs free energy of adsorption ($\Delta$G$_{ads}$) for Gdm$^+$. We find a notable similarity between the $\Delta$G$_{ads}$ values for Gdm$^+$ and SCN$^-$ at the air-water interface. Simulations are used to clarify the molecular origins of this driving force for adsorption and to provide further structural details on the solvation of Gdm$^+$ at the air-water interface. \par

\section*{Results and Discussion}
\textbf{Bulk Absorption spectroscopy}. The bulk UV absorption spectrum of GdmCl (black trace) and NaCl (green trace) in water are shown in Figure \ref{fig:fig.1}. The charge-transfer-to-solvent (CTTS) transition of Cl$^-$ has been previously characterized, with a peak centered near ca. 180 nm \cite{Blandamer1970Theory,Marin2019Ultraviolet}. Antol et al. theoretically calculated the UV spectrum of Gdm$^+$ and found a strongly absorbing $\pi$-$\pi$* transition in the DUV region \cite{Antol2014Guanidine}. Pointwise DUV-SHG measurements taken at various wavelengths for 3 M GdmCl solutions were normalized to pure water and plotted relative to the maximum intensity (black squares) in Figure \ref{fig:fig.1}. A steady increase in SHG signal is seen from 220 – 200 nm, matching the low energy shoulder of the bulk absorption spectrum. It is difficult to generate a precise interfacial spectrum via SHG due to the nature of the technique, which relies on single wavelength measurements.\par
Many of the anions studied by DUV-SHG have strong CTTS transitions with large molar absorption coefficients ($\varepsilon$) in the 10$^2$ - 10$^4$ M$^{-1}cm^{-1}$ range \cite{Blandamer1970Theory}, making them ideal candidates for resonant signal enhancement. SHG studies have shown that these transitions may shift by ca. 5 - 20 nm at the interface relative to the bulk, a result of the sensitivity of CTTS transitions to a chromophore’s local solvation environment \cite{Petersen2004Confirmation,Petersen2005Enhanced,Petersen2004Direct,Devlin2022On}. At 200 nm, the molar absorption coefficient of Gdm$^+$ was much larger than that of Cl$^-$, viz. $\varepsilon$ = 375 vs 59 M$^{-1}cm^{-1}$, respectively. Previous DUV-SHG studies reported a low average SHG response for pure NaCl solutions across the 200 – 225 nm wavelength range \cite{Petersen2006Probing}. This indicates that we should expect negligible resonant signal enhancement at these wavelengths due to the Cl$^-$ CTTS transition. The following DUV-SHG experiments were conducted with a 400 nm input (200 nm SHG) wavelength, where resonant signal enhancement contributions are primarily due to the Gdm$^+$ $\pi$-$\pi$* transition. \par

\begin{Figure}   
    \centering
    \includegraphics[width=\linewidth]{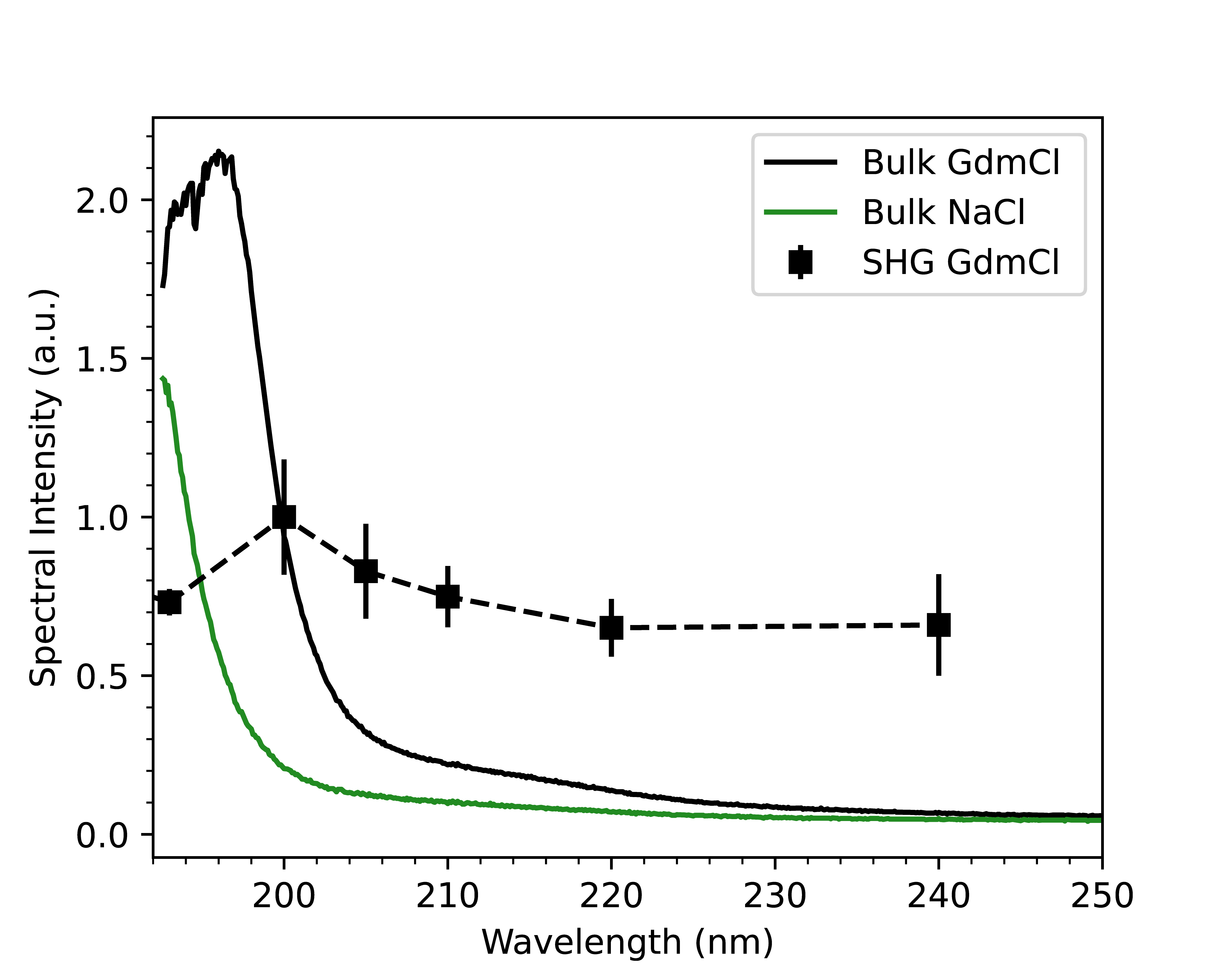} 
    \captionof{figure}{DUV absorption spectra of bulk GdmCl (black trace) and NaCl (green trace) solutions. The $\pi-\pi$* transition of Gdm$^+$ is clearly visible along with the onset of the Cl$^-$ charge-transfer-to-solvent (CTTS) transition. SHG response of 3 M GdmCl (black squares), normalized to pure water at varying SHG wavelengths, is shown joined with a dashed line serving as a guide to the eye. }
    \label{fig:fig.1}
\end{Figure}
\textbf{Langmuir Adsorption Model}. A Langmuir adsorption model is used to extract the Gibbs free energy of adsorption from concentration-dependent DUV-SHG intensities. Although modified adsorption models are not uncommon in literature for fitting SHG intensities \cite{Ng2022Iron(III),Eckenrode2005Adsorption,Castro1991Energetics}, a Langmuir model proves to be the simplest and fits the data presented here well. A thorough discussion of the Langmuir model used in SHG studies can be found elsewhere \cite{Petersen2006Probing,Petersen2004Confirmation}; here only a brief description is provided. \par
DUV-SHG is a second-order nonlinear optical process and under the electric dipole approximation, signal from the centrosymmetric bulk environment is forbidden and quadrupole and magnetic dipole contributions are omitted \cite{Shen1989Optical,Heinz1991Chapter}. Thus, the signal generated from DUV-SHG arises from the few outermost molecular layers where inversion symmetry is broken, typically expected to be 1 nm depth. By tuning the input energy so that the generated second harmonic field is resonant with an electronic transition of a chromophore, we gain signal enhancement and directly probe the number of interfacial chromophores. The intensity of the second harmonic signal ($I_{2\omega}$) is proportional to the second order nonlinear susceptibility of the interfacial species ($\chi _{water}^{(2)}$, $\chi _{Gdm^+}^{(2)}$) and the squared intensity of the driving field ($I_\omega^2$):
\[I_{2\omega} \propto |\chi _{water}^{(2)} + \chi _{Gdm^+}^{(2)}|^2  \times I_\omega^2 \tag{1}\]

Nonlinear susceptibilities become complex quantities when resonant with a transition. At the frequencies used here, $\chi _{water}^{(2)}$ is non-resonant and real, but $\chi _{Gdm^+}^{(2)}$ is resonant and therefore a complex quantity with both real and imaginary components.  Nonlinear susceptibilities can be described as an ensemble of molecular hyperpolarizabilities, i.e. the degree to which each molecule responds to a driving electric field,
\[ \chi _{Gdm^+}^{(2)} = \sum_{i} \beta _i = N_s \times \langle \beta \rangle _{orient.} \tag{2} \]
which is further denoted as the product between the number of molecules ($N_s$) and their orientationally averaged hyperpolarizability ($\langle \beta \rangle$$_{orient.}$). It is assumed that the magnitude of $\langle \beta \rangle_{orient.}$ remains constant during the experiment and therefore that the orientations of the interfacial molecules remain unperturbed by ion-ion interactions as the bulk concentration changes. Vibrational SFG studies on the orientation of SCN$^-$ found that the linear molecule remains tilted at  \ang{44} from surface normal at the air-water interface between a wide range of molar concentrations \cite{Hao2020Specific}. Similarly, simulations show that Gdm$^+$ preferentially adsorbs to the interface in a single orientation with minimal fluctuation, and is independent of bulk concentration \cite{Wernersson2011Orientational,Heiles2015Hydration}. Thus, changes in the SHG signal become directly proportional to the number of surface molecules adsorbing to the interface, and this permits the use of a Langmuir adsorption model,
\begin{multline}
\frac{I_{2\omega}}{I_\omega^2} = ( \textbf{A} + \textbf{B}\frac{X_{Gdm^+}}{(1-X_{Gdm^+})e^{\frac{\Delta G}{RT}}+ X_{Gdm^+}})^2\\ + (\textbf{C}\frac{X_{Gdm^+}}{(1-X_{Gdm^+})e^{\frac{\Delta G}{RT}}+ X_{Gdm^+}})^2 \tag{3} 
\end{multline}
where \textbf{A} represents the real water susceptibility, and \textbf{B} and \textbf{C} are the real and imaginary contributions to the Gdm$^+$ susceptibility, respectively. SHG intensities are fit using Eq.3 and allowing \textbf{A}, \textbf{B}, and \textbf{C} to vary when extracting the Gibbs free energy of adsorption ($\Delta G_{ads}$). A full derivation of Eq. 3 and further discussion on the assumptions made by incorporating a Langmuir model can be found in the SI Appendix. \par
\textbf{DUV-SHG}. SHG intensities of NaCl solutions normalized to pure water are plotted against bulk concentration in Fig. 2. The pH of NaCl solutions was adjusted to the pH of GdmCl solutions at each respective concentration via HCl. A weak linear response in SHG intensity with electrolyte concentration is seen here. This response has been previously attributed to an interfacial thickening of the water layer \cite{Bian2008Increased}. As the bulk concentration of the solution increases, interfacial water molecules rearrange, effecting a change in their hyperpolarizability resulting in a linear increase in the SHG intensity, normalized to neat water. Attempts to fit the data in Fig. 2 to Eq. 3 were unsuccessful as it does not conform to a typical adsorption pattern. Although we expect to be off-resonance with the Cl$^-$ CTTS transition, ions adsorbing to the interface should still elicit changes in the SHG response as the interfacial concentration increases, as noted in other non-resonant SHG studies \cite{Bian2008Increased,Rehl2019New}. Since we observe a predominantly flat response from pH-adjusted NaCl solutions, we suspect that the chloride anion resides beyond the probe depth of our DUV-SHG measurements.\par
\begin{Figure}   
    \centering
    \includegraphics[width=\linewidth]{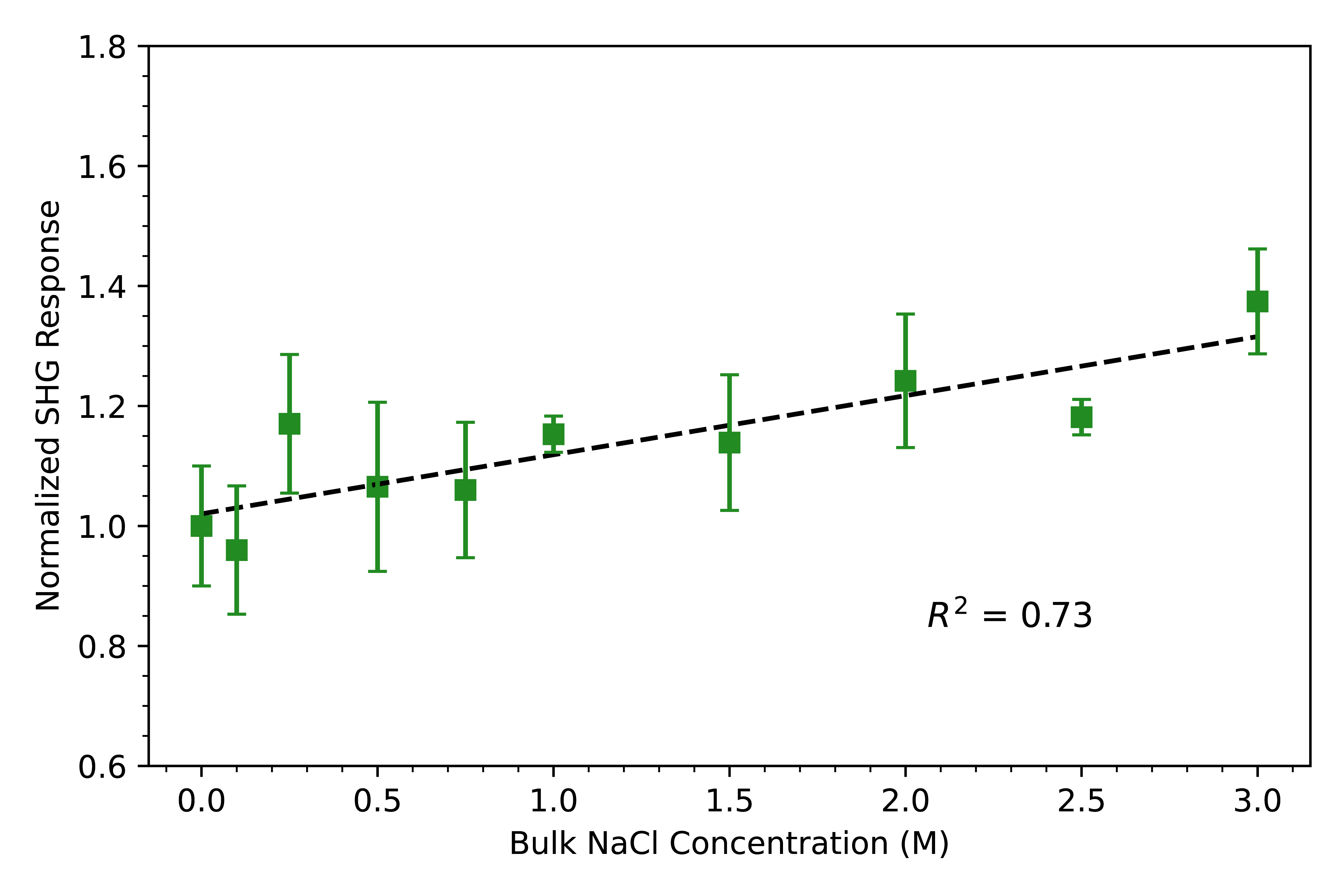} 
    \captionof{figure}{SHG response ($\omega_{SHG}$ = 200 nm) of pH-adjusted NaCl solutions normalized to pure water plotted against bulk concentration. The dashed line represents a linear fit with an $R^2$ = 0.73 and serves as a guide to the eye. A Langmuir fit to these data was attempted and can be found in S1. }
    \label{fig:fig.2}
\end{Figure}
The results in Figure \ref{fig:fig.2} are unsurprising, since it has been demonstrated before that the chloride anion exhibits a weak surface-affinity for the air-water interface compared to larger halides. This is evidenced by density profile plots of Cl$^-$ lacking a large population increase at the interface \cite{Jungwirth2002Ions} and by SHG studies reporting a weak response from pure NaCl solutions \cite{Petersen2006Probing}. Recently, Seki et al. reported heterodyne-detected SFG measurements that quantified a salting-out effect that Cl$^-$ induces on co-solvated anions such as SCN$^-$ \cite{Seki2023Ions}. Their work finds a maximum 50\% increase in the surface population of known surface-active anions after the addition of NaCl. This effect is attributed to Cl$^-$ exhibiting hydrophilic character in solution, thereby residing mainly in the bulk and promoting a greater number of hydrophobic anions to the interface.\par
Figure \ref{fig:fig.3}.A shows the SHG intensities of GdmCl solutions normalized to the pure water response and plotted with respect to bulk concentration. There is a clear signal dependance on concentration that was not seen for NaCl solutions, which we attribute to the presence of Gdm$^+$ at the interface. For comparison, we have also reexamined the prototypical SCN$^-$ anion via DUV-SHG in Figure \ref{fig:fig.3}.B. The maximum normalized SHG response is relatively low for Gdm$^+$ compared to SCN$^-$. Analyzing the SHG response in Figure \ref{fig:fig.1}, the signal at the resonant wavelength is only 1/3 greater than the off-resonance signal, likely due to the $\pi - \pi$* transition contributing minimal signal enhancement. Spectral shifts in the resonant transition can affect the SHG intensity; however, as noted above, there is no obvious shift in the peak Gdm$^+$ SHG response compared to the bulk absorption.\par
\begin{Figure} 
    \centering
    \includegraphics[width=\linewidth]{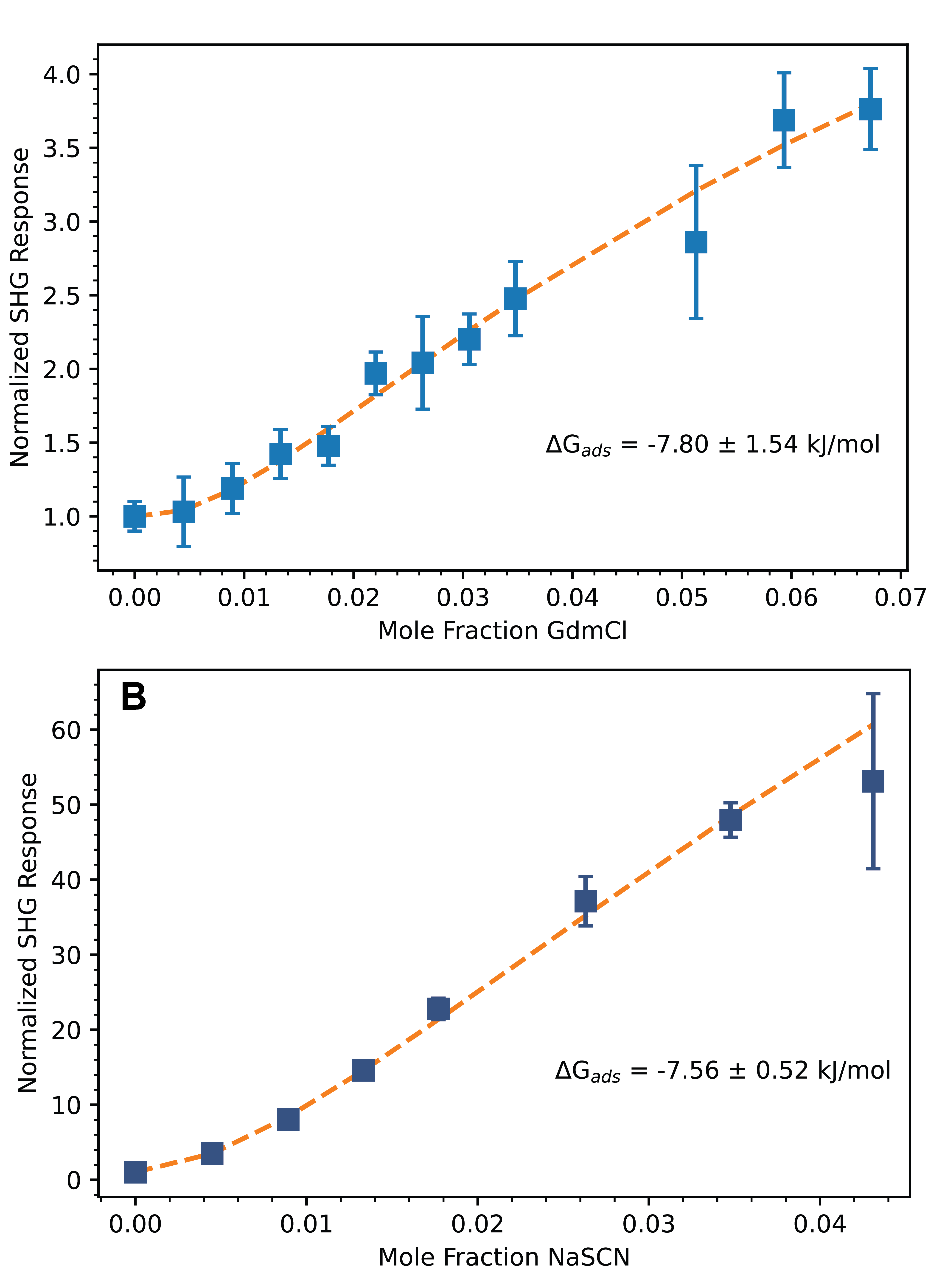} 
    \captionof{figure}{Comparison of interfacial adsorption for cations and anions at the air-water interface. (A) Normalized DUV-SHG response ($\omega _{SHG}$ = 200 nm) of GdmCl solutions at the air-water interface (blue squares) plotted against bulk concentration. Data are fit to a Langmuir adsorption model (orange dashed line) with an extracted Gibbs free energy of -7.8 ± 1.54 kJ/mol. (B) Normalized SHG intensity ($\omega _{SHG}$ = 200 nm) for NaSCN solutions at the air-water interface (dark-blue squares) plotted against bulk concentration. Data are fit to a Langmuir adsorption model (orange dashed line) with an extracted Gibbs free energy of -7.56 ± 0.52 kJ/mol. Uncertainties reported are one $\sigma$}
    \label{fig:fig.3}
\end{Figure}
Investigating the parameters generated from fitting SHG intensities to Eq.3, we find that both systems share similar \textbf{A} values, as expected, since it is due to the real water response. However, the \textbf{B} and \textbf{C} parameters, which are due to the ion resonances, differ between the two systems, with both parameters being lower for Gdm$^+$. Examining the magnitude of the susceptibility of both ions ($\sqrt{\textbf{B}^2 + \textbf{C}^2}$), we find the net $\chi _{Gdm^+}^{(2)}$ to be 82\% less than $\chi _{SCN^-}^{(2)}$, as reflected in the SHG intensities plotted in Figure \ref{fig:fig.3}. We recognize that these fit parameters were extracted for studies at a single resonant wavelength and that a more rigorous approach in determining the resonant and non-resonant ion contributions involves additional concentration dependent measurements at off-resonant wavelengths. Given the already low resonant SHG response of Gdm$^+$, tracking the off-resonance response would prove challenging, given our current sensitivity, and would likely be better suited to heterodyne-detected methods. Irrespective of this, we can clearly discern changes in the resonant SHG signal with GdmCl concentration, hence our fit to a Langmuir model strongly-supports previous studies concluding that Gdm$^+$ resides at the air-water interface \cite{Ekholm2018Anomalous,Werner2014Surface}.\par
Despite the order of magnitude difference in SHG oscillator strength between both systems, the extracted $\Delta G_{ads}$ for both ions are surprisingly within error. The Gibbs free energy of adsorption for SCN$^-$ at hydrophobe-water interfaces has been extensively measured by DUV-SHG experiments and is well-reproduced here for the air-water interface using a 400 nm incident wavelength ($\Delta G_{ads}$ = -7.56 ± 0.52 kJ/mol). For Gdm$^+$ ions, this study produced $\Delta G_{ads}$ = -7.80 ± 1.54 kJ/mol, indicating that both ions share a similar propensity for adsorbing to the interface. These ions are two well-studied protein denaturants that are situated on the far-chaotropic end of the Hofmeister series \cite{Okur2017Beyond}, and therefore might be expected to exhibit similar ion effects and properties in water. Litman et al. recently reported strong perturbations in the vibrational SFG water O-H signal with increasing perchlorate concentration, an effect they expect to extend to SCN$^-$ and other chaotropic ions \cite{Litman2024Surface}. The recently reevaluated hydration free energy of Gdm$^+$ is -78.4 kcal/mol, a value 8.4 kcal/mol larger than that for SCN$^-$ \cite{Neal2020Molecular,Benjamin2022Structure}. In solution, both monovalent ions achieve stability through charge delocalization, engendering ‘soft’ ion character. Given the similarities between these ions, it begs the question: Would the properties of each ion, including the $\Delta G_{ads}$, differ if both ions were co-solvated in water? Balos et al. examined the interaction of Gdm$^+$ and SCN$^-$ with a model amide, and observed competing interactions between the two ions in solution \cite{Balos2017Anionic}. The overall ion-amide interaction rotational effects were non-additive for GdmSCN, implying that individual SCN$^-$-amide and Gdm$^+$-amide interactions were not compounded when both ions were present in solution, indicating a contest between both ions for the amide. Signal from DUV-SHG measurements of GdmSCN solutions would certainly be dominated by SCN$^-$ ions and lack a straightforward route to deconvoluting the respective signal contributions.\par
\textbf{Molecular Dynamics}. Molecular dynamics simulations were performed with the LAMMPS software package\cite{Thompson2022LAMMPS}. All simulations were performed in an ensemble with fixed number of particles, temperature, and volume, using a Langevin thermostat at 300 K and a relaxation time of 200 fs, and in a periodic 18.5 \AA {} $\times$ 18.5 \AA {} $\times$ 130 \AA {} simulation cell. All simulations contained 450 water molecules and a variable number of GdmCl ion pairs in a slab geometry with a large vapor region in the z direction. GdmCl ions were modeled using a previously parametrized force field \cite{England2011Role}. To prevent the trivial global translation of the slab in the $z$ direction, the center of mass is subject to a harmonic constraint at $z_{CoM} = 0$. Water was modeled using the non-polarizable tip4p/2005 model and the polarizable swm4-ndp model which employs Drude oscillators to explicitly capture water polarizability. Furthermore, to determine whether an explicit description of water polarizability was needed, simulations using the electronic continuum correction (ECC) were also performed, which scale the charge of ions in water by a factor of 0.7 to account for excess charge screening due to water polarizability. Drude oscillator simulations were performed using the LAMMPS DRUDE package, with a dual Langevin thermostat fixing the temperature of fictitious Drude particles to 1K with a relaxation time of 100 fs and a timestep of 1 fs. Symmetrized Drude force fields were used to improve numerical stability\cite{Dodin2023Symmetrized}. Non-polarizable simulations were performed with a timestep of 2 fs. Water molecules were held rigid using the symplectic rigid body integrator in the LAMMPS RIGID package\cite{Kamberaj2005Time}.\par
Simulations of 2.5 M GdmCl solutions contained 20 GdmCl ion pairs in 450 water molecules. Potentials of mean force (PMF’s) in the dilute limit are constructed using an Umbrella Sampling scheme \cite{Kästner2011Umbrella} in which the Gdm$^+$ ions are subject to a series of harmonic bias potential $U_{bias}=\frac{k}{2}(z-z_0)^2$. Drude oscillator simulations were performed using 51 Umbrella windows, spaced 0.5 \AA {} apart in the interval 0 \AA {} $< z <$ 25 \AA {} with spring constants of 10 kCal/mol/\AA$^2$. Non-polarizable simulations were performed using 18 windows spaced 1 \AA {} apart in the interval 0 \AA {} $< z <$ 18 \AA {} with $k$ = 2.5 kCal/mol/\AA$^2$, and 9 windows spaced 0.5 Å apart in the interval 18 \AA {} $< z <$ 22 \AA {} with $k$ = 1- kCal/mol/\AA$^2$. All umbrella simulations were equilibrated for 500 ps with a production run of 4 ns. Simulations of the concentrated simulations were unbiased and were equilibrated for 5 ns with a total production run of 50 ns. In all cases, the Gibbs dividing surface was at approximately $z$ = 20 \AA. Biased simulations were then reweighted into the unbiased ensemble using the Grossfield implementation \cite{Grossfield2002WHAM:} of the Weighted Histogram Analysis method\cite{Kumar1995Multidimensional,Kumar1992weighted,Souaille2001Extension} with 50 equally spaced bins in the range 0 \AA {} $< z <$ 23 \AA {} for non-polarizable simulations and the range 0 \AA $< z <$ 25 \AA {} for Drude oscillator simulations.\par
\begin{figure*}[t] 
    \centering
    \includegraphics[width=.5\linewidth]{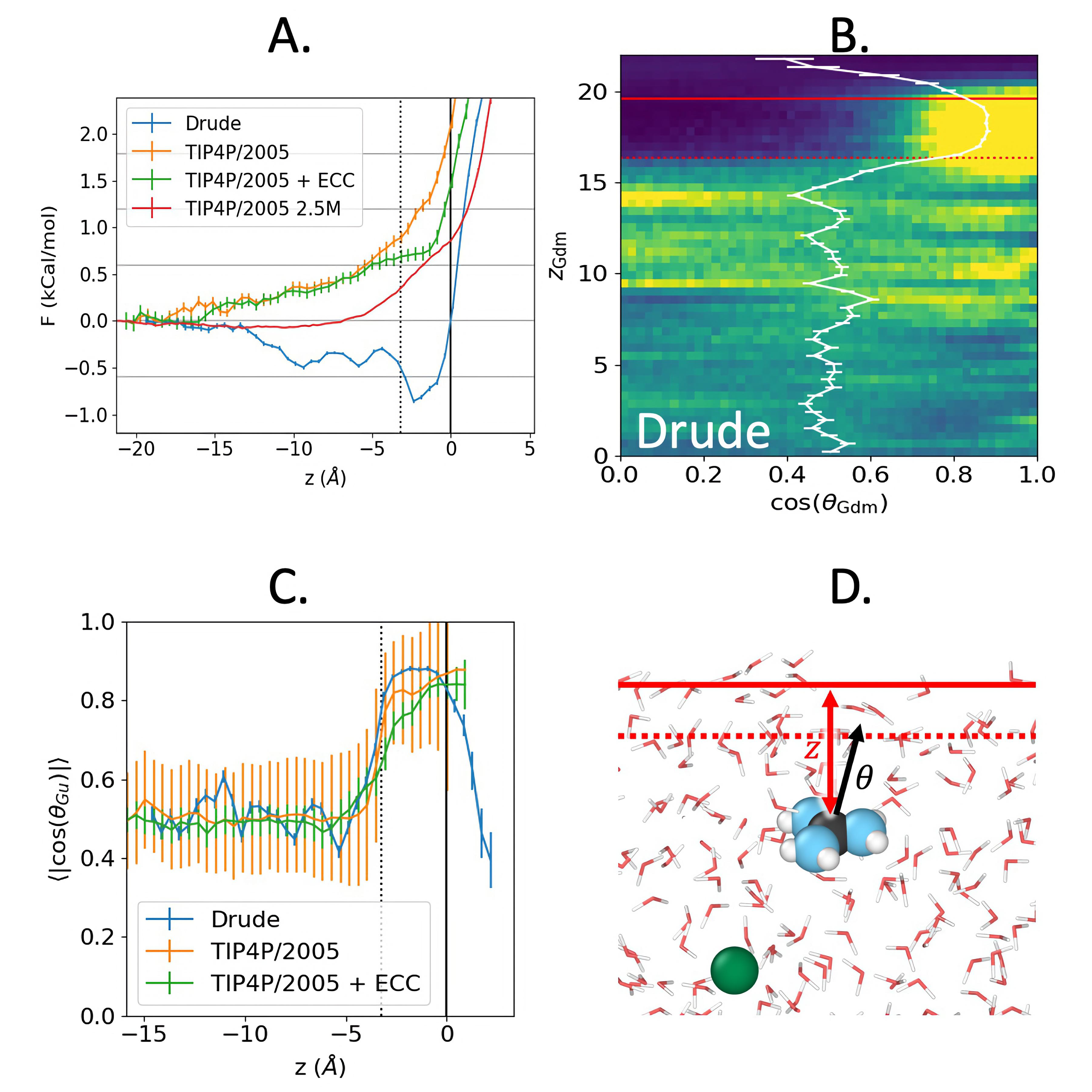} 
    \captionof{figure}{Molecular dynamics simulations of GdmCl with varying water models. (A) Potentials of mean force of Gdm$^+$ ions in the dilute limit for three models for water polarizability, and in a concentrated 2.5 M solution for the tip4p/2005 water model. Horizontal lines show kT = 2.48 kJ/mol. (B) Orientation histogram of Gdm$^+$ in explicitly polarizable water as a function of distance from the Gibbs dividing surface (solid red line). The dashed red trace indicates the boundary of the subsurface layer. The white trace shows the average orientation as a function of $z$. (C) Average orientation of Gdm$^+$ cation as a function of distance from the interface for 3 different water models in the dilute limit. (D) MD Snapshot showing the definition of $z$ and $\theta$. In all figures the Gibbs dividing surface and bound of the subsurface layer are shown in the solid and dashed lines respectively.}
    \label{fig:fig.4}
\end{figure*}

The potential of mean force (PMF) of Gdm$^+$ in the dilute limit relevant to the Langmuir isotherm is shown in Figure \ref{fig:fig.4}.A as a function of distance from the average interface (i.e. the Gibbs dividing surface where the density falls to half its bulk value). We perform this computation for three water models that account for the polarizability of water at different levels – the non-polarizable tip4p/2005 water model \cite{Abascal2005general}, an implicitly polarizable model based on the Electronic Continuum Correction (ECC) that accounts for water’s polarizability at the mean-field level by scaling down ionic charges \cite{Pegado2012Solvation}, and the explicitly polarizable Drude swm4-ndp model of water which includes dynamically changing dipoles on each water molecule \cite{Lamoureux2003Modeling,Lamoureux2006polarizable,Lemkul2016Empirical}. In stark contrast to our spectroscopic measurements, non-polarizable tip4p and implicitly polarizable ECC water models predict that the Gdm$^+$ is repelled more than a nanometer away from the interface. Accounting for the explicit polarizability of water using the Drude model reverses this trend, predicting an interfacial enhancement of Gdm, consistent with experiment, and corresponding to $\Delta G_{ads}$ = -3.59 ± 0.17 kJ/mol.  A previous study of GdmCl in a non-polarizable water model similarly predicted the repulsion of Gdm+ from the liquid-vapor interface, and proposed that any surface enhancement of Gdm$^+$ must be attributed to specific orientations of the ion near the interface \cite{Wernersson2011Orientational}. Our studies suggest an alternative interpretation, wherein the enhanced interfacial affinity of Gdm$^+$ can be attributed to the polarizability of interfacial water. Water molecules near the interface have smaller dipole moments than those in the bulk, facilitating the displacement of interfacial solvent to the bulk that must occur when an ion approaches the interface. Specifically, if a Gdm$^+$ cation is moved from the bulk region to the interface, the molecules that formerly occupied its new interfacial location move to fill its initial location of the bulk. In moving from the interface to the bulk, their dipole moments increase relative to that of bulk water, increasing the attractive interactions they experience with other water molecules, providing a thermodynamic driving force pushing the ion towards the interface that is absent in non-polarizable water models.  In previous studies of the absorption mechanism for thiocyanate, this effect was termed “solvent repartitioning” \cite{Otten2012Elucidating}.\par
While our Drude simulations are qualitatively consistent with spectroscopic measurements, they estimate a $\Delta G_{ads}$ that is a factor of 2 smaller than experiment. The experimental measurements are performed at much higher concentrations than our simulations. In Figure \ref{fig:fig.4}.A we also compare the dilute PMF’s assumed in the Langmuir model, to those of Gdm$^+$ in a 2.5 M solution of GdmCl, corresponding to a mole fraction of 0.0425, in
\end{multicols}

\begin{multicols*}{2}
\noindent the same regime as experimental data. In the concentrated system, we see a significant stabilization, $\sim$ 1 kT relative to the dilute limit. This change in PMF indicates that ions in GdmCl do not approach the interface independently, as assumed by the Langmuir model, but that ion-ion correlations can play a significant role in their interfacial activity. This observation is consistent with recent studies which predicted the surprising formation of like-charge Gdm$^+$-Gdm$^+$ ion pairs in bulk solution \cite{Vazdar2018Arginine,Shih2013Cation-cation} that may contribute to interfacial ion-ion correlations.\par
We also calculate how the orientational distribution of the planar Gdm$^+$ cation changes near the interface’s broken symmetry.  In Figure \ref{fig:fig.4}.B, we consider the joint probability distribution of the angle between the normal vector to the Gdm$^+$ plane and the normal of the interface, and its distance to the average interface, for the Drude water model. Similarly, Figure \ref{fig:fig.4}.C shows the average orientation of the Gdm$^+$ ion as it approaches the interface for the 3 water models studied. In all three cases, we see that a strong orientational bias emerges in the subsurface layer, where the ions lie coplanar with the interface. In this subsurface layer, the cation remains fully solvated, with a single water molecule between it and the gas phase. In contrast, to the PMF where the repulsion or attraction to the interface extends for over a nanometer, the orientational bias is strongly localized within the subsurface layer.\par
The interface-specific DUV-SHG measurements are therefore selectively probing a highly localized subsurface population of interfacial co-planar Gdm$^+$ cations. Wernersson et al, reported an orientational dependance on the surface adsorption of Gdm$^+$, similarly finding an interfacial preference for parallel orientations of the cation relative to the surface boundary \cite{Wernersson2011Orientational}. This orientational preference was rationalized by considering the weakly hydrated faces of the planar Gdm$^+$ and the location of its hydrogen bonding N-H groups in the molecular plane. At the interface, where molecules tend to be poorly solvated relative to the bulk, the most energetically favorable configuration of Gdm$^+$ is that wherein its face is normal to the air-water boundary \cite{Heiles2015Hydration}. Bulk density profiles for Gdm$^+$ in solution evidence only a small percentage of cations at the interface, a result of the strong interfacial orientational preference excluding all but a single Gdm$^+$ orientation \cite{Wernersson2011Orientational}. Since $\chi _{Gdm^+}^{(2)}$ is dependent on the interfacial density of ions and their hyperpolarizabilities, we speculate that there is a probable low interfacial concentration of Gdm$^+$ at the interface contributing to the signal intensity. This reasoning also fits with surface tension measurements of GdmCl solutions, which indicate that ions should be net depleted from the interface, and indeed only a small subpopulation of Gdm$^+$ demonstrates an affinity for the interface \cite{Breslow1990Surface}. Thus, the $\Delta G_{ads}$ extracted from experiment is a result of probing only a small percentage of specifically oriented Gdm$^+$ that at high concentrations exhibit ion-ion correlations. The lack of quantitative agreement between computational and experimental predictions of $\Delta G_{ads}$ may then arise due to a breakdown in the assumptions of the Langmuir model for correlated electrolytes.\par

\textbf{Sample Preparation}. All glassware was left overnight in a bath of Alnocromix (Alconox Inc) and concentrated sulfuric acid (Sigma Aldrich) to remove organic debris, then rinsed with ultrapure 18.2 M$\Omega$ {} water (Millipore MilliQ) before use. The salts GdmCl(Sigma Aldrich, $>$ 98\% purity), NaCl (Sigma Aldrich, $>$ 99\% purity), and NaSCN (Sigma Aldrich, $>$ 98\% purity) were used as is with ultrapure water to make stock solutions. NaCl solutions were pH adjusted using reagent grade HCl (Sigma Aldrich). Aliquots of each solution were drawn from the bottom of stocks using borosilicate serological pipettes to avoid organics on the surface and deposited into Petri dishes immediately before measurement. \par
\textbf{Experimental Design}. The laser setup has been described previously. Briefly, the output of a Ti-Sapphire amplifier (Spectra Physics Spitfire Ace, 2 mJ, 1 kHz, 100 fs) centered at 800 nm is directed through a type-I beta barium borate (Edmund Optics, $\beta$-BBO) crystal to produce the fundamental 400 nm beam. For measurements at fundamental wavelengths below and above 400 nm, the 800 nm output is directed into an optical parametric amplifier (TOPAS Prime). Input power is modulated by a rotating circular variable neutral density filter operating at 1 Hz and input polarization is controlled by a half-wave plate and polarizer. A 10 cm lens focuses the p-polarized fundamental beam onto liquid samples at an angle of \ang{60} relative to the sample surface normal. The colinearly reflected fundamental and p-polarized SHG signal are collimated with a 10 cm lens then separated spectrally in series by a dichroic mirror, Pellin-Broca prism, and monochromator (Acton, SpectraPro 2150i). A solar-blind photomultipler tube (Hamamatsu, R7154PHA) and boxcar integrator (Stanford Research Systems, SR250) serve as the photon counting detector. All sample measurements are normalized to the SHG response of pure water samples taken daily between sample measurements to account for day-to-day fluctuations of the setup. Bulk UV-Vis measurements were made using a Shimadzu UV-2600 spectrometer. 
\subsection*{Conclusion}
Using DUV-SHG spectroscopy, we provide the first direct experimental evidence for a polyatomic cation adsorbing to the air-water interface. By fitting concentration-dependent SHG data to a simple Langmuir adsorption model, we have determined the Gibbs free energy of adsorption for Gdm$^+$ ions to be -7.80 ± 1.54 kJ/mol (one sigma uncertainty). We highlight that this ion appears to adsorb to the interface just as strongly as does the prototypical chaotropic anion, SCN$^-$. The exact underlying mechanistic details of Gdm$^+$ interfacial adsorption remain to be determined, but future studies will undoubtably provide new insight that may even extend to intriguing Gdm$^+$-protein interactions. Our previous work elucidated the mechanism by which SCN$^-$ adsorbs to the air-water interface. We found ions partitioning to the interface to be enthalpically driven via displacement of interfacial water molecules into the bulk, where more favorable hydrogen bonding occurs due to increased coordination, but entropically impeded due to suppression of interfacial capillary waves \cite{Otten2012Elucidating}. It is not unreasonable to suggest that Gdm$^+$ may follow a similar mechanism, but we cannot be certain without further theory and experiments to uncover additional features of Gdm$^+$ ion adsorption ($\Delta H_{ads} , \Delta S_{ads}$). Considering that experiments have shown SCN$^-$  adsorbing to both oil-water and graphene-water interfaces with similar $\Delta G_{ads}$ to have underlying mechanistic differences \cite{Devlin2022On}, and new studies highlighting the likelihood of like-charge contact ion-paring of Gdm$^+$ ions in solution \cite{Vazdar2018Arginine,Shih2013Cation-cation},  we clearly do not yet have a complete description of Gdm$^+$  behavior at the air-water interface.  Nevertheless, this work clearly illustrates that cations can also exhibit a strong propensity for the air-water interface.    
\subsection*{Acknowledgments}
This work was supported by the Department of Energy, Office of Basic Energy Sciences, through the Chemical Sciences Division at the Lawrence Berkeley National Laboratory under contract \#CH403503 and the Condensed Phase and Interfacial Molecular Science Program (CPIMS), in the Chemical Sciences Geosciences and Biosciences Division of the Office of Basic Energy Sciences of the U.S. Department of Energy under Contract No. DE-AC02-05CH11231.\par
\noindent Corresponding Authors:\par
\noindent *dlimmer@berkeley.edu
*saykally@berkeley.edu
\par
\subsection*{Data Availability}
Data and code for this work are available online through Zenodo.
https://zenodo.org/doi/10.5281/zenodo.13381652

\end{multicols*}
\begin{multicols}{2}

\bibliography{Gdm}{}
\end{multicols}


\section*{Supplementary material}

# Adsorption of Guanidinium Cations to the Air-Water Interface

Franky Bernal[1,2], Amro Dodin[2], Constantine Kyprianou[1], David T. Limmer[1,2,3,4,*], Richard J. Saykally[1,2,*]

[1]Department of Chemistry, University of California, Berkeley, CA 94720, USA
[2]Chemical Sciences Division, Lawrence Berkeley National Laboratory, Berkeley, CA 94720, USA
[3]Materials Science Division, Lawrence Berkeley National Laboratory, Berkeley, California 94720, USA
[4]Kavli Energy NanoScience Institute, Berkeley, California 94720, USA

**Langmuir Isotherm Derivation**

The use of a Langmuir adsorption model for analyzing SHG data has been covered extensively(1, 2). The SHG intensity can be expressed as being proportional to the second order nonlinear susceptibility of the interfacial water and guanidinium ions (Gdm$^+$) ($\chi^{(2)}_{water}, \chi^{(2)}_{Gdm^+}$) and the square of the fundamental intensity ($I_\omega^2$).

$$I_{2\omega} \propto \left| \chi^{(2)}_{water} + \chi^{(2)}_{Gdm^+} \right|^2 \times I_\omega^2 \quad (1)$$

Second-order nonlinear susceptibilities can be described as the product of the number of interfacial species ($N$) and their averaged hyperpolarizabilities ($\langle \beta \rangle_{orient.}$).

$$\chi^{(2)}_{ion} = \sum_i \beta_i = N \times \langle \beta \rangle_{orient.} \quad (2)$$

Combining Eq. 1 and 2, we can express the ratio of SHG intensity and fundamental intensity squared as follows:

$$\frac{I_{2\omega}}{I_\omega^2} = |N_{water} \times \langle \beta \rangle_{water} + N_{Gdm^+} \times \langle \beta \rangle_{Gdm^+}|^2 \quad (3)$$

Guanidinium has a resonant π-π* transition in the UV and therefore its hyperpolarizability is complex with a real and imaginary signal component. The water molecules are far from an electronic transition and therefore contribute a purely real signal. This modifies Eq 3 to the following:

$$\frac{I_{2\omega}}{I_\omega^2} = (N_{water} \times \langle \beta \rangle_{water} + N_{Gdm^+} \times Re\{\langle \beta \rangle_{Gdm^+}\})^2 + (N_{Gdm^+} \times Im\{\langle \beta \rangle_{Gdm^+}\})^2 \quad (4)$$

Dividing each term by the number of water molecules leads to:

$$\frac{I_{2\omega}}{I_\omega^2} = \left( \langle \beta \rangle_{water} + \frac{N_{Gdm^+}}{N_{water}} \times Re\{\langle \beta \rangle_{Gdm^+}\} \right)^2 + \left( \frac{N_{Gdm^+}}{N_{water}} \times Im\{\langle \beta \rangle_{Gdm^+}\} \right)^2 \quad (5)$$

We now simplify the equation to:

$$\frac{I_{2\omega}}{I_\omega^2} = (A + N_s \times B)^2 + (N_s \times C)^2 \quad (6)$$

Here, the $N_S$ term is a concentration of interfacial ions. The A, B, and C parameters are the real water, real ion, and imaginary ion signal contributions, respectively. Following a Langmuir adsorption model, we assume that ions solvated in the bulk can exchange positions with a water molecule at the surface.

$$water_{surf} + Gdm^+_{bulk} \leftrightarrow water_{bulk} + Gdm^+_{surf} \quad (7)$$

We use this relation to generate an equilibrium expression for the surface adsorption of Gdm+ ions and ultimately find an expression for $N_S$:

$$K_{ads} = \frac{[water]_{bulk} \times [Gdm^+]_{surf}}{[water]_{surf} \times [Gdm^+]_{bulk}} \quad (8)$$

By assuming there is a maximum number of surface sites and only one ion can occupy a given site, we solve for the concentration of surface ions:

$$N_S = [Gdm^+]_{surf} = [sites]_{max} \times \frac{[Gdm^+]_{bulk}}{[water]_{bulk} \times K_{ads}^{-1} + [Gdm^+]_{bulk}} \quad (9)$$

We substitute Eq. 9 into 6, convert concentrations to mole fraction ($X_{Gdm+}$), and expand the equilibrium constant to include the Gibbs free energy ($K_{ads} = e^{\frac{-\Delta G}{RT}}$). The maximum surface sites term is incorporated into the B and C parameters. The final fitting expression is used to extract Gibbs free energies of adsorption for Gdm$^+$ ions at the interface:

$$\frac{I_{2\omega}}{I_\omega^2} = \left(A + B \frac{X_{Gdm+}}{(1-X_{Gdm+})e^{\frac{\Delta G}{RT}} + X_{Gdm+}}\right)^2 + \left(C \frac{X_{Gdm+}}{(1-X_{Gdm+})e^{\frac{\Delta G}{RT}} + X_{Gdm+}}\right)^2 \quad (10)$$

**Fit Parameters**

Guanidinium fit parameters were obtained from a Levenberg-Marquardt algorithm for Eq. 10.

| | |
|---|---|
| A | 0.99966700 ± 0.04947297 |
| B | -0.15702257 ± 0.88749779 |
| C | 2.70369596 ± 0.33748800 |
| G | -7801.55761 ± 1542.27109 |
| T | 293 (fixed) |
| R | 8.314 (fixed) |

Guanidinium parameter correlations

| |
|---|
| C(B, G) = 0.905 |
| C(B, C) = 0.373 |
| C(A, B) = -0.503 |
| C(A, C) = 0.114 |
| C(A, G) = -0.251 |
| C(C, G) = 0.716 |

NaSCN fit parameters

| | |
|---|---|
| A | 1.00271539 ± 0.04983243 |
| B | 3.36592869 ± 2.80799798 |
| C | 14.5960181 ± 1.28740369 |
| G | -7560.49827 ± 515.716266 |
| T | 293 (fixed) |
| R | 8.314 (fixed) |

NaSCN parameter correlations

| | |
|---|---|
| C(B, G) = 0.905 | |
| C(B, C) = 0.600 | |
| C(A, B) = -0.408 | |
| C(A, C) = 0.127 | |
| C(A, G) = -0.125 | |
| C(C, G) = 0.873 | |

As referenced in the main text, the SHG response of NaCl solutions did not fit well to a Langmuir model. The attemped fit is shown below.

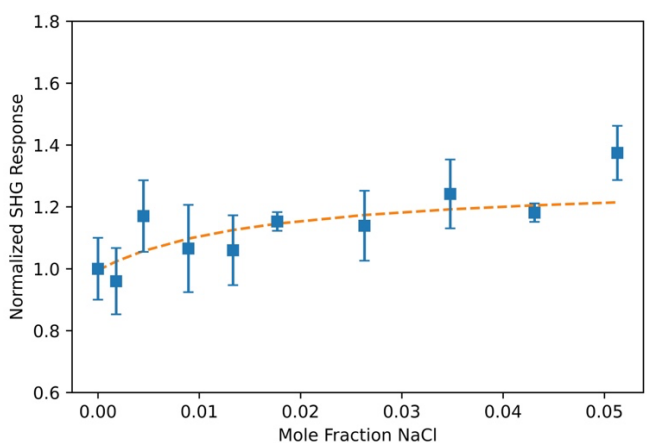

**Figure S1**. Attempted Langmuir fit to NaCl SHG data. Fit results in unphysical parameters and Gibbs free energy of adsorption.

NaCl Fit parameters

| | |
|---|---|
| A | 0.99747304 ± 0.04753699 |
| B | 0.13367770 ± 9.75178727 |
| C | 1.5796e-04 ± 68837.1960 |

| G | -10226.6738 ± 167125.548 |
|---|---|
| T | 293 (fixed) |
| R | 8.314 (fixed) |

NaCl parameter correlations

| |
|---|
| C(B, G) = +0.9998 |
| C(B, C) = -1.0000 |
| C(A, B) = -0.5202 |
| C(A, C) = +0.5183 |
| C(A, G) = -0.5067 |
| C(C, G) = -0.9998 |


**Deep UV Second Harmonic Generation Spectroscopy**

DUV-SHG experiments are carried out under two-photon resonant conditions according to the expression below:

$$\chi^{(2)} \propto \sum_n \frac{|\mu_{0,n}(\alpha_{0,n})|}{\omega_n - \omega_{SHG} - i\Gamma} \quad (11)$$

Here, the interfacial susceptibility is proportional to the one-photon transition dipole matrix ($\mu_{0,n}$) and two-photon absorption polarizability tensor element ($\alpha_{0,n}$) of the resonant ion. $\omega_n$ and $\omega_{SHG}$ are the frequencies of the resonant transition and generated SHG photon. $\Gamma$ is the linewidth of the resonant transition. SHG signal enhancement occurs when the transition is both one- and two-photon active.



\end{document}